\documentclass[12pt, twoside, a4paper]{report}
 \usepackage{setspace}
 \pdfoutput=1
\usepackage[dvips]{graphicx}
\usepackage{latexsym}
\usepackage{epsf}
\usepackage{epsfig}
\usepackage{color}
\usepackage{a4}
\usepackage{float}
\usepackage{amsmath} 
\usepackage{amssymb}
\usepackage{amsxtra}
\usepackage{fancyhdr}
\usepackage{hyperref}
\hypersetup{
    linktoc=all,     
    linkcolor=red,  
    pdfstartview={XYZ null null 1.00},
}

\begin{document}
\begin{titlepage}
\begin{center}
\topskip 0pt
\vspace*{5cm}
\Large{QUINTESSENTIAL INFLATION IN MIMETIC DARK MATTER}
\vspace*{0.5cm} 
\end{center}
\begin{center}
\normalsize {Ali Rida Khalifeh}
\end{center}
\vspace*{0.5cm}
\begin{center}
\normalsize {\textit{Physics Department, American University of Beirut, Lebanon}}
\end{center}
\vspace*{3cm}
A quintessential Inflation (QI) scenario from Mimetic Dark Matter (MDM) is presented in this paper. This scenario, which is based on the MDM model presented by Chamseddine and Mukhanov \cite{chams}, uses a potential that is defined on three time intervals:the first during inflation, the second $10^{-65}s$ after the end of inflation and the third after that. The resulting energy density of the universe is constant during inflation, followed by that of a matter/radiation dominated  universe, and finally ends with a constant energy density corresponding to dark energy. The scale factor has an accelerating expansion nature during and after inflation. It will be shown how this is still a viable scenario, even if the scale factor after inflation is not that of a decelerating De Sitter universe.
\end{titlepage}

\chapter{Introduction}
Quintessential Inflation models have been introduced drastically in the literature in an attempt to link inflation to the later stages of the universe's evolution\cite{peebles}-\cite{model4}.  The key element in this unification is the fact that both inflaton, the field describing inflation, and quintessence, the field describing dark energy, are both dynamical scalar fields that are describing an accelerating expansion of the Universe. However, one cannot avoid to mention that some models have used gauge fields to describe these phenomena \cite{antonino}, nevertheless we are not going to tackle on these in this paper. The way this is usually done is by adding the Lagrangian of a scalar field to that of general relativity, while assuming an FRW metric: $	ds^2=dt^2-a(t)^2\delta_{ij}dx^idx^j$ where $a(t)$ is the scale factor and $\delta_{ij}$ is the Kronecker delta. This will result in the following action:
\begin{equation}\label{1.1}
S=\int d^4x \sqrt{-g}(-\dfrac{1}{2}R+\frac{1}{2}\dot{\phi}^2+V(\phi))
\end{equation}
where g is the determinant of the metric, R is the Ricci scalar, $\phi$ is the QI field, the dot denotes derivative with respect to time and V($\phi$) is the potential which describes the dynamics of the field. Instead of introducing scalar fields from outside into the Lagrangian, Chamseddine and Mukhanov wrote the physical metric in the following way\cite{chams}:
\begin{equation}\label{1.2}
g_{\mu\nu}=\tilde{g}_{\mu\nu}(\tilde{g}^{\alpha\beta}\partial_{\alpha}\phi\partial_{\beta}\phi)\equiv P\tilde{g}_{\mu\nu}
\end{equation}
where $\tilde{g}_{\mu\nu}$ is an auxiliary metric, $\phi$ is (for the moment) a random scalar field and $\partial_{\alpha}$ denotes partial derivative with respect to $x^{\alpha}$. In this way, one might say that the conformal mode of the metric has been isolated, for the physical metric is invariant under a conformal transformation of the auxiliary metric. Moreover, one can see from [\ref{1.2}]that such an isolation results in the following constraint equation:
\begin{equation}\label{1.3}
g^{\mu\nu}\partial_{\mu}\phi\partial_{\nu}\phi=1
\end{equation}
which will be essential in specifying the scalar field later on. From here, the resulting action would be:
\begin{equation}\label{1.4}
S= -\frac{1}{2}\int{d^{4}x \sqrt{-g(\tilde{g}_{\mu\nu},\phi)}[R(g_{\mu\nu}(\tilde{g}_{\mu\nu},\phi))+L_m]}
\end{equation}
where $L_m$ is the matter content of the system. Thus, the gravitational field acquires an extra longitudinal degree of freedom, in addition to the two transverse ones representing the graviton. This extra degree of freedom will prove to be useful. Now, the resulting equation of motion (upon varying the action with respect to the metric) gives the Einstein tensor $G^{\mu\nu}$ in terms of the stress-energy tensor $T^{\mu\nu}$ and an extra term $\tilde{T}^{\mu\nu}$:
\begin{equation}\label{1.7}
G^{\mu\nu}=T^{\mu\nu}+\tilde{T}^{\mu\nu}
\end{equation} where 
\begin{equation}\label{1.8}
\tilde{T}^{\mu\nu}=(G-T)g^{\mu\alpha}g^{\nu\beta}\partial_{\alpha}\phi\partial_{\beta}\phi
\end{equation}
 with $G$ and $T$ being the trace of the Einstein tensor and the stress-energy tensor, respectively. Now compare [\ref{1.8}] to that of a perfect fluid, $T^{\mu\nu}=(\rho+p)u^{\mu}u^{\nu}-pg^{\mu\nu}$ where $\rho$ is the energy density, $p$ is the pressure and $u^{\mu}$ is the 4-velocity which satisfies the normalization $u^{\mu}u_{\mu}=1$. By doing the following identifications: $p=0$, $\rho\equiv G-T$ and $u^{\mu}\equiv g^{\mu\alpha}\partial_{\alpha}\phi$, one can see that this extra degree of freedom imitates ``dust".
 Therefore, one is not obliged to introduce any type of matter from outside to explain the phenomena attributed to dark matter, rather now one has to extract hidden fields from the metric to explain these phenomena. Choosing the FRW metric, one can see that by solving [\ref{1.3}], the resulting scalar field would be: 
 \begin{equation}\label{1.9}
 \phi=t
 \end{equation}
 which will be used through out this paper.
 \newline
  Furthermore, such isolation can be introduced into the action in the following way \cite{golovnev}-\cite{cosmology mimetic}:
\begin{equation}\label{1.10}
S=\int d^4x\sqrt{-g}[-\dfrac{1}{2}R+\lambda(g^{\mu\nu}\partial_{\mu}\phi\partial_{\nu}\phi-1) -V(\phi)+L_m]
\end{equation}
where $\lambda$ is a Lagrange multiplier. Now, varying the action with respect to the metric, we get the ``modified" Einstein equation:
 \begin{equation}\label{1.11}
G_{\mu\nu}-T_{\mu\nu}-2\lambda\partial_{\mu}\phi\partial_{\nu}\phi-g_{\mu\nu}V=0
\end{equation} Taking the trace of [\ref{1.11}], we get the following relation for $\lambda$: 
\begin{equation}\label{1.12}
\lambda=\dfrac{1}{2}(G-T-4V)
\end{equation} which means, after plugging back in \ref{1.11}, 
\begin{equation}\label{1.13}
G_{\mu\nu}=(G-T-4V)\partial_{\mu}\phi\partial_{\nu}\phi+g_{\mu\nu}V(\phi)+T_{\mu\nu}
\end{equation}
As was done above, the first two terms on the right hand side of [\ref{1.13}] can be identified with the stress energy tensor of a perfect fluid with pressure:
\begin{equation}\label{1.14}
\tilde{p}=-V
\end{equation} and energy density, \begin{equation}\label{1.15}
\tilde{\rho}=G-T-3V
\end{equation}
The time-time component of [\ref{1.13}] gives the Friedman equation:
\begin{equation}\label{1.16}
H^2=\dfrac{1}{3}\tilde{\epsilon}=\dfrac{1}{a^3}\int a^2Vda
\end{equation}
Multiplying [\ref{1.16}] by $a^3$, and differentiating with respect to time, while substituting $y=a(t)^{3/2}$, we get:

\begin{equation}\label{1.17}
\ddot{y}-\dfrac{3}{4}V(t)y=0
\end{equation}
(see \cite{cosmology mimetic} for detailed derivation).
\newline
From here, the work of this paper is based on choosing an appropriate potential for QI, plug it in [\ref{1.17}] and study the resulting expansion of the Universe and get its energy density. We will end up with a conclusion and future works.

\chapter{The Model}
In this chapter, we will consider a quintessential inflation model for MDM. The structure of the potential was inspired from a paper by Peebles and Vilenkin \cite{model4}, although the exact details are not the same. Moreover, the content of the potential is based on the dynamics of a slow rolling field, but using MDM. We will then go onto considering the appropriate potential in MDM that would produce almost the same effect of QI. 
\section{Inspiration from Slow Rolling Cosmology }
Consider a slow-rolling scalar field, with the following potential: 
\begin{equation}\label{2.1}
V=e^{-\alpha \phi}
\end{equation} with $\phi=\ln(t)$ \cite{peebles} is the QI field and $\alpha$ is a constant; thus we are using a power law potential. The scale factor that comes from a slowly rolling field (using non-modified General Relativity)is :
 \begin{equation}\label{2.2}
a=a_0 exp(\dfrac{\alpha}{3(2-\alpha)}t^{2-\alpha}).
\end{equation} with $a_0$ being a constant of integration. The energy density of such a field would be: 
\begin{equation}\label{2.3}
\rho=\dfrac{1}{3M_{pl}^2}(\dfrac{\alpha}{t^{\alpha-1}})^2
\end{equation} where $M_{pl}$ is the Planck mass. With an appropriate choice of $\alpha$, this model shows an exponential expansion of the universe, but with an energy density that goes like $t^{-2}$. This energy density is that of radiation and matter\cite{mukhanov}. On the other hand, if we take the following potential: 
\begin{equation}\label{2.4}
V=\beta e^{-\phi}
\end{equation} the scale factor then is: \begin{equation}\label{2.5}
a=a_0exp(\dfrac{1}{3}\beta t)
\end{equation} with the same definition for $\phi=\ln(t)$. Moreover, the energy density is now: \begin{equation}\label{2.6}
\rho=\dfrac{1}{3M_{pl}^2}\beta^2
\end{equation} which is a constant.  From here, we see that to produce an energy density that represents matter-radiation dominated universe (i.e goes like $t^{-2}$) directly after inflation, and reaches an asymptote, the potential must be a combination between the two. Combining the two potentials together, while substituting the form of $\phi$, we get a potential of the form:
\begin{equation}\label{2.7}
V=At^{-\alpha}+Bt
\end{equation} So, let's try to see what physics will be produced from MDM if we use a polynomial potential, with [\ref{1.9}] as the scalar field. Moreover, Peebles and Vilenkin used the following potential:
 \begin{equation}\label{2.8}
 V=
 \begin{cases}
 \lambda(\phi^4+M^4); \qquad \phi<0 \\ \\
 \dfrac{\lambda M^8}{\phi^4+M^4}; \qquad \phi\geq 0
 \end{cases}
 \end{equation}
where $M$ is a parameter to be fixed by data. Although there are some differences between this work and the one in \cite{model4}, it would be interesting to see what could such a potential (that is one which is defined on two intervals of $\phi$, or equivalently two time intervals) produce when used in MDM, and it will be shown that some important results appear.

\section{The Potential}
Let's use the following potential in MDM: \begin{equation}\label{2.9}
V=\dfrac{2\alpha}{3}(1-\alpha)(t-t_0)^{-\alpha}+\dfrac{1}{3}[\alpha(t-t_0)^{-\alpha}+\beta]^2
\end{equation} The origin was shifted to $t_0$, the time at which inflation ends. As compared to [\ref{2.7}], this has been done in order to make sure that the dynamics are centered at the end of inflation . Moreover, the choice of the coefficients is made in such a way that no clustering of constants occurs later on. Plugging this potential in [\ref{1.9}], we get: \begin{equation}\label{2.10}
\ddot{y}-[\dfrac{\alpha(1-\alpha)}{2}(t-t_0)^{-\alpha}-\dfrac{1}{4}[\alpha(t-t_0)^{1-\alpha}+\beta]^2]y=0
\end{equation} the solution of this equation will give us the scale factor to be\cite{kimke}: \begin{equation}\label{2.11}
a=a_0exp[\dfrac{\alpha}{3(2-\alpha)}(t-t_0)^{2-\alpha}+\dfrac{1}{3}\beta(t-t_0)]
\end{equation} and an energy density for the mimetic matter: 
\begin{equation}\label{2.12}
\tilde{\rho}=\dfrac{1}{3M_{pl}^2}[\dfrac{\alpha}{(t-t_0)^{\alpha-1}}+\beta]^2
\end{equation} One can see that if $\alpha$ is very small, $\tilde{\rho}\propto t^{-2}$ at the beginning, that is near the end of inflation, and then as $t\rightarrow \infty$, $\tilde{\rho}\rightarrow \dfrac{1}{3M_{pl}^2}\beta^2$. So far, what we have is exactly the behavior we expect. What remains is fixing the parameters $\alpha$ and $\beta$ to produce the desired measurable quantities. However, there's still something wrong with this potential. First, the energy density and the scale factor might diverge, unless we have a good choice of the parameter $\alpha$ at the boundaries. Second, if $t<t_0$, and we have a fractional power in the energy density and the scale factor, we will get imaginary numbers. This is something definitely we don't want in real measurable quantities.
\\  Therefore the solution will be as follows: we will separate the potential into two parts, one before inflation ($t\leq t_0$) and the other after inflation ($t> t_0$). We will then match these two values at $t=t_0$. This way, we will have the term $t_0-t$ during inflation ($t\leq t_0$) and the term ($t-t_0$) after inflation. By doing this, we have solved the issue of having imaginary numbers. Now, concerning the divergence issue, we look at the form of the scale factor and the energy density in [\ref{2.10}] and [\ref{2.11}]. To avoid divergences, during inflation, at $t=t_0$, $2-\alpha$ must be positive, so must be $1-\alpha$. Therefore the solution to avoid divergence at $t=t_0$ as we approach it from the left, is to have: \begin{equation*}
\alpha<1
\end{equation*} Now, for $t>t_0$, keeping the same form of the potential, our concern is at $\infty$, since there we don't want the energy density to diverge, rather we want it to be a very small number. Moreover, the scale factor should not diverge at $t=t_0$. Therefore, $2-\alpha'>0$ and $\alpha'-1>0$(we are using $\alpha'$ just to distinguish it from the constant during inflation). So for the post-inflation phase: \begin{equation*}
1<\alpha'<2
\end{equation*} The final result for the potential that would produce a quintessential inflation model in MDM is: \begin{equation}\label{2.13}
V=
\begin{cases}
\dfrac{2\epsilon}{3}(1-\epsilon)(t-t_0)^{-\epsilon}+\dfrac{1}{3}[\epsilon(t-t_0)^{-\epsilon}-\beta']^2, & \qquad t> t_0 \\ \\
\dfrac{2(2-\epsilon)}{3}(\epsilon-1)(t_0-t)^{\epsilon-2}+\dfrac{1}{3}[(2-\epsilon)(t_0-t)^{\epsilon-2}+\beta]^2, & \qquad t\leq t_0
\end{cases}
\end{equation} and the corresponding scale factor is : \begin{equation}\label{2.14}
a=
\begin{cases}
a_0exp[\dfrac{2-\epsilon}{3\epsilon}(t-t_0)^{\epsilon}-\dfrac{1}{2}\beta'(t-t_0)], & \qquad t> t_0 \\ \\
a_0exp[\dfrac{\epsilon}{3(2-\epsilon)}(t_0-t)^{2-\epsilon}+\dfrac{1}{2}\beta(t_0-t)], & \qquad t\leq t_0
\end{cases}
\end{equation} while the energy density: \begin{equation}\label{2.15}
\rho=
\begin{cases}
\dfrac{1}{3M_{pl}^2}[\dfrac{2-\epsilon}{(t-t_0)^{1-\epsilon}}-\beta']^2 & \qquad t> t_0 \\ \\
\dfrac{1}{3M_{pl}^2}[\dfrac{\epsilon}{(t_0-t)^{\epsilon-1}}+\beta]^2 & \qquad t\leq t_0
\end{cases}
\end{equation} where $\epsilon$ is an infinitesimal number. Now, to determine $\beta$, we have to use the number of e-folds of inflation. If inflation is to last for 70 e-folds, then: \begin{equation}\label{2.16}
N=\int_{t_i}^{t_0} Hdt\equiv 70
\end{equation} where $t_i$ is the time at which inflation is supposed to have started. According to the model first presented by Guth, inflation should start at $t_i=10^{-36}$ and end at $t_0=10^{-32}$ \cite{guth}. Plugging in these numbers into [\ref{2.15}], we get: 
\begin{equation}\label{2.17}
\beta\approx 7\times 10^{32}
\end{equation} On the other hand, $\beta'$ is determined by matching the value of the energy density at infinity to that of the cosmological constant \cite{constant}. This will result in \begin{equation}\label{2.18}
\beta'\approx \sqrt(3) \times 10^{-23}
\end{equation}
Before continuing into checking the results of the model, there's one last issue that needs to be tackled. It is apparent from the form of the energy density in [\ref{2.15}] that it diverges at $t=t_0$. This might mean that there's a discontinuity in the energy density at the end of inflation. We can approximately solve this issue by looking at how much time it takes $\rho$ to go from $\infty$ to the value at $t=t_0$ if we are approaching it from the left (i.e. using the expression of the energy density for $t\leq t_0$). If we plug in the value of $\beta$ in $\rho$ for $t\leq t_0$, we get the energy density at $t=t_0$ to be of the order of $10^{100}$. Setting this value to be that of the field for $t>t_0$, and calculating the time interval, it turns out that it takes the energy density approximately $10^{-65}s$ to go from $\infty$ to $10^{100}$. Since it is a very short period of time, we can insert yet a third interval, directly after inflation, which extends for only $10^{-65}s$. During this time, the energy density, the scale factor and the potential take approximately constant values corresponding to those at the end of inflation. Therefore the third part (now is the second)will end $10^{-65}s$ after the end of inflation and there will be no divergence in the form of $\rho$ when taken in $t>t_0+10^{-65}$, and the continuity problem is therefore solved. This maneuver wont affect any previous or later calculations, since it extends over an extremely short period of time.
 These equations will result in the plots below for the scale factor and the energy density (we have used an $\epsilon=0.01$).\begin{figure}
\centering
\includegraphics[width=0.7\linewidth]{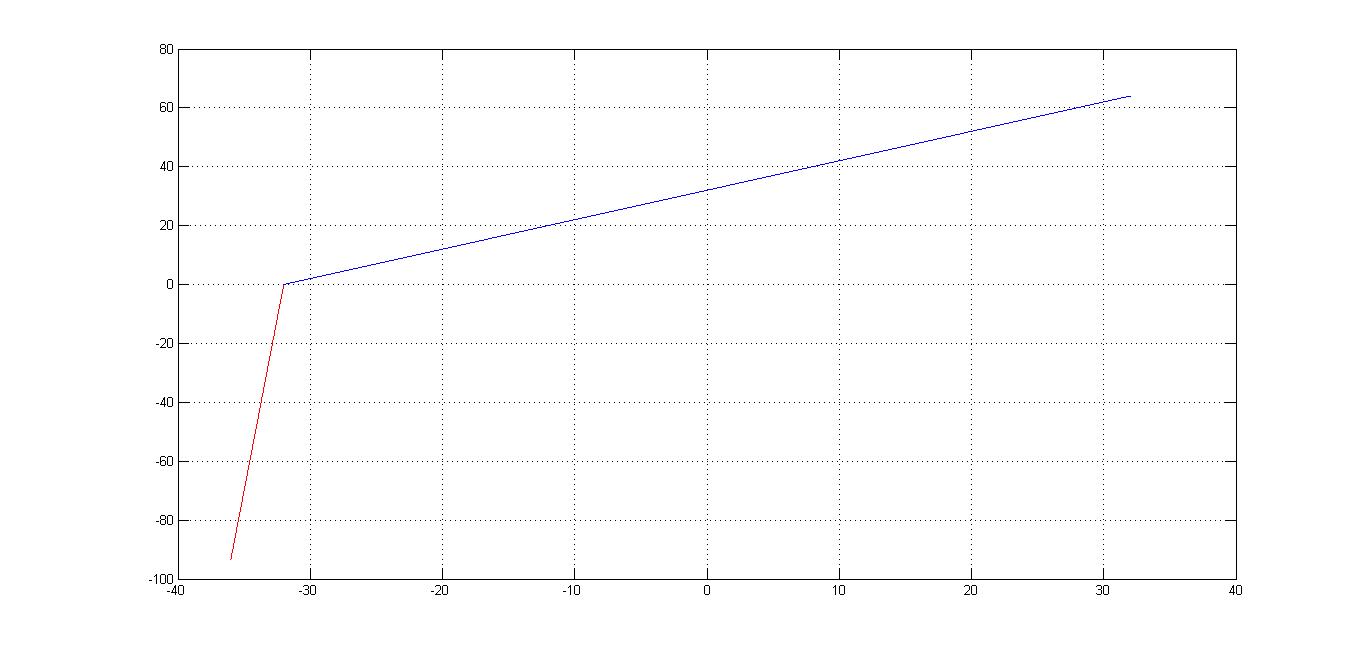}
\caption{Plot of the logarithm of the scale factor as a function of the logarithm of time, for the two regimes: during inflation (red) and after inflation(blue). It is clear that there's a huge expansion in the universe during inflation, and a moderate one afterwards. The value of $\epsilon$ that has been used is 0.01}
\label{fig:abeforeandafter}
\end{figure}
\begin{figure}
\centering
\includegraphics[width=0.7\linewidth]{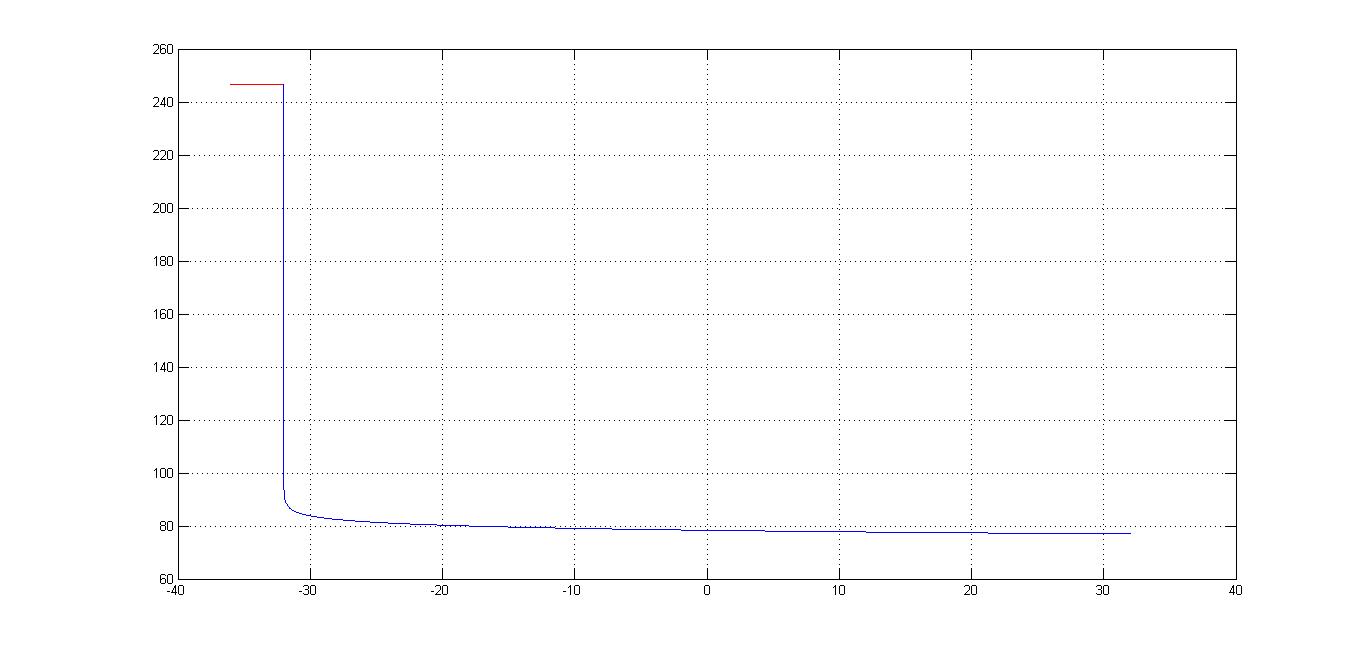}
\caption{Plot of the logarithm of the energy density as a function of the logarithm of time for the two regimes}
\label{fig:epsbeforeandafter2}
\end{figure}
 From the first plot, it is clear that the scale factor is increasing with $\ddot{a}>0$, which implies it is an accelerated expansion of the universe. The expansion during inflation is much steeper than that after it, which is exactly what's needed, since the universe cannot keep on accelerating at the same rate as that during inflation. Furthermore, concerning the energy density plot, the graph shows a constant energy density during inflation, which is a characteristic of inflation. In addition, the energy density reaches an asymptote as $t\rightarrow \infty$, which is nothing but Quintessence. The inflation parameters, with the above expressions of the scale factor, are consistent with the conditions for inflation: \begin{equation}\label{2.19}
 \epsilon=-\dfrac{\dot{H}}{H^2}=\propto 10^{-35}<<1; \qquad \eta=\dfrac{d\ln\epsilon}{dN}\sim 0
 \end{equation}
 \\
 \\ Now we will discuss what type of perturbations does this model lead to. As was pointed out in \cite{cosmology mimetic}, in order to have a difference between the short wavelength perturbations and the long ones, a term of the form $\dfrac{1}{2}\gamma(g^{\mu\nu}\nabla_{\mu}\nabla_{\nu}\phi)^2$ ($\gamma$ is a constant)must be added to the action. Moreover, the short wavelength perturbations have been shown to be independent of the choice of the potential. On the other hand, for long wavelength perturbations, we do have a dependence on the choice of the potential, for it depends on the scale factor. The integral in the fluctuation is taken over the period of inflation, that is from $t_i=10^{-36}$ to $t_0= 10^{-32}$. At the end of inflation, the two terms in $exp$ of [\ref{2.14}] die away. Therefore we can say that the integral is dominated by the lower limit,  and since we have an exponential expansion, we can approximate the form of the scale factor to be $a\sim \exp(\beta (t-t_0))$. This will make the integral much easier to calculate. From here, we get the perturbations in the scalar field to be: \begin{equation}\label{2.20}
 \delta\phi=A\dfrac{1}{a}\int a^2d\eta=\dfrac{A}{\beta}\simeq\dfrac{1}{H}
 \end{equation} which corresponds to perturbations in an inflationary stage \cite{cosmology mimetic}. To get the factor A, we have to match the value of the short wavelength perturbations to that of the long wavelength. The result is:
  \begin{equation}\label{2.21}
 A\sim\sqrt{\dfrac{c_s}{\gamma}}\dfrac{H}{k^{3/2}}
 \end{equation} with H being evaluated at $\eta\sim \dfrac{1}{c_sk}$, $c_s$ and $k$ are the speed of sound and the wave number, respectively. These results are in agreement with \cite{cosmology mimetic}

\chapter{Conclusion}
In this paper, a Quintessential inflation scenario from the Mimetic Dark Matter model \cite{chams} has been presented. The potential used to produce such a scenario is defined on three time intervals, one during inflation($t=10^{-36}-10^{-32} s$), one after inflation that extends for $10^{-65}s$ and the third after that. The parameters of the potential were set in a way to produce 70 e-folds inflation and to have an energy density corresponding to the one measured today, representing Dark Energy. The scale factor after inflation is that of an accelerating universe, in contrast to a decelerating De-Sitter Universe as it has been presented in the literature \cite{mukhanov}.
\newline
 However, in the non-modified General theory of Relativity, an energy density of matter/radiation is accompanied by a decelerating Universe, this is why usually it is required to have a decelerating universe after inflation. But in this case, from the equations of MDM, one obtains an energy density of matter/radiation dominated universe. The important thing is the energy density rather than the scale factor. Since in this model we got the required energy density but not the "usual" scale factor, we can avoid the problem of explaining why the universe should decelerate after inflation. Rather, at the end of inflation, the universe looses enough energy for it to remain accelerating, but with a slower rate than the one during inflation. Moreover, concerning the structure of the potential, this form has been presented in physical problems other than Cosmology, mainly in electrostatics \cite{jackson}, in order to produce physically acceptable and non-diverging electric fields. Any discontinuities or divergences in the potential should not be considered dangerous, as long as the Physical quantities are smooth and well behaved. Of course we still have to check whether this will still result in the required Nucleosynthesis and we have to see if the temperature perturbations that arise matches those of the CMB. These will be handled in future work.

\bibliographystyle{ieeetr}

\end{document}